\documentclass[twocolumn,amssymb,amsmath,aps,prb,showpacs,10pt]{revtex4-2}
\usepackage[dvips]{graphicx}
\usepackage{color}
\usepackage{multirow}
\usepackage[hidelinks]{hyperref} 
\usepackage{amsmath}
\usepackage{amssymb}
\usepackage{soul}
\hypersetup{
	colorlinks   = true, 
	urlcolor     = blue, 
	linkcolor    = blue, 
	citecolor    = blue 
}

\begin{document}
\title {Does carrier localization affect the anomalous Hall effect?}

\author{Prasanta Chowdhury$^1$, Mohamad Numan$^1$, Shuvankar Gupta$^2$,  Souvik Chatterjee$^3$, Saurav Giri$^1$,  Subham Majumdar$^1$}
\email{sspsm2@iacs.res.in}

\affiliation{$^1$School of Physical Sciences, Indian Association for the Cultivation of Science, 2A \& B Raja S. C. Mullick Road, Jadavpur, Kolkata 700 032, India}

\affiliation{$^2$Condensed Matter Physics Division, Saha Institute of Nuclear Physics, 1/AF, Bidhannagar, Kolkata, 700 064, India}

\affiliation{$^3$UGC-DAE Consortium for Scientific Research, Kolkata Centre, Sector III, LB-8, Salt Lake, Kolkata 700106, India}

\begin{abstract}	
The effect of carrier localization due to electron-electron interaction in anomalous Hall effect is elusive and there are contradictory results in the literature. To address the issue, we report here the  detailed transport study including the Hall measurements on  $\beta$-Mn type  cubic compound Co$_7$Zn$_7$Mn$_6$ with chiral crystal structure, which  lacks global  mirror symmetry. The alloy orders magnetically below $T_c$ = 204 K, and reported to show spin glass state at low temperature. The longitudinal resistivity  ($\rho_{xx}$) shows a pronounced upturn below $T_{min}$ = 75 K, which is found to be associated with carrier localization due to quantum interference effect.  The upturn in $\rho_{xx}$  shows a  $T^{1/2}$ dependence and it is practically insensitive to the externally applied magnetic field, which indicate  that electron-electron interaction is primarily responsible for the low-$T$ upturn. The studied sample shows considerable value of anomalous Hall effect below $T_c$. We found that  the localization effect is present in the  ordinary Hall coefficient ($R_0$),  but we failed to observe any signature of  localization in the anomalous Hall resistivity or conductivity. The absence of localization effect in the anomalous Hall effect in Co$_7$Zn$_7$Mn$_6$ may be due to large carrier density, and it warrants further theoretical investigations, particularly with systems having broken mirror symmetry. 
	
\end{abstract}

\maketitle

\section{Introduction}
The ordinary Hall (OH) effect, which  arises from the deflection of the moving charge carriers due to Lorentz force, was discovered in 1879~\cite{OH}, and its underlying mechanism is in general considered to be well comprehended. In contrast, the anomalous Hall effect(AHE)~\cite{AHE1,AHE2} observed in magnetic materials has remained relatively subtle. The phenomenon is intriguing both from fundamental point of view as well as for its potential applications in sensors, memories and logics~\cite{Application1,Dirty}. It is now well recognized that there are mainly three mechanisms responsible for AHE, namely, intrinsic mechanism, skew scattering and side-jump~\cite{Dirty}. Karplus and Luttinger were first to suggest that intrinsic mechanism arises from transverse velocity of the Bloch electrons induced by spin-orbit interaction (SOI) and interband mixing~\cite{KL}, and recently Xiao {\it et al.} reinterpreted it in terms of Berry curvature of the occupied Bloch states~\cite{Berry}. The other two mechanisms (skew scattering and side-jump) are  extrinsic in nature and they arise from the asymmetric scattering of the conduction electrons by the impurities in presence of  SOI as proposed by Smit~\cite{smit1,smit2} and Berger~\cite{Berger}. Depending upon  its origin, the anomalous Hall resistivity ($\rho_{xy}^{AHE}$)  scales  differently with longitudinal resistivity ($\rho_{xx}$).  The  skew scattering is generally observed in highly conducting metals ($\rho_{xx} \lesssim$ 10$^{-6}$~$\Omega$ cm) with low amount of impurities~\cite{Dirty} and it varies as $\rho_{xy}^{AHE}\propto\rho_{xx}$. On the other hand,  both intrinsic and side-jump mechanisms follow $\rho_{xy}^{AHE}\propto\rho_{xx}^2$.
\par
However, the situation is far more complex, if a comparatively  higher degree of disorder is present in the magnetic metal ($\rho_{xx} \gtrsim$ 10$^{-3}~\Omega$ cm).  In presence of disorder, the quantum effects become more prominent, and the system can show localization of charge carriers due to  electron-electron coulomb interaction (EEI), disordered induced weak localization (WL), or Kondo effect~\cite{Dirty,WL,RMP}. As a result, metallic $\rho_{xx}(T)$ exhibits an upturn ($d\rho_{xx}(T)/dT <$ 0)  at low temperature ($T$), and  a resistivity minimum (at $T_{min}$) is observed ~\cite{RMP,SNKAUL,WL1,Mitra,localized}. The effect of  quantum corrections has been extensively studied for longitudinal conductivity and conventional Hall effect ~\cite{RMP,WL,EEI3}, but it is still poorly understood in case of AHE.  
\par
There are few recent theoretical and experimental works addressing the effect of localization on AHE~\cite{EEI1,EEI2,WLC1,Ni,FePt,Mitra,CoFeB,HgCrSe,FeP1,localized,Hg_Theory,ZrVCoSn}. It is shown theoretically that WL does not contribute towards side-jump mechanism, but it can have nonzero contribution in skew scattering~\cite{WLC1,EEI1}. On the other hand, EEI correction to AHE identically vanishes for both skew scattering and side-jump due to general symmetry reasons~\cite{EEI1,EEI2}. The above theoretical prediction of the absence of  EEI correction towards AHE  was experimentally verified in Co$_2$FeSi Heusler alloy thin film ~\cite{SNKAUL} and Zr$_{1-x}$V$_x$Co$_{1.6}$Sn semimetal~\cite{ZrVCoSn}. On the other hand, WL correction was experimentally observed in polycrystalline Fe, Ni, FePt and amorphous CoFeB films~\cite{FeP1,localized,Ni,FePt,CoFeB}. However, there are  some disordered systems, which do not follow the above rules. For example, WL effect in AHE  is found to be absent in the disordered ferromagnets Ga$_{1-x}$Mn$_x$As ~\cite{Mitra}. Similarly, a pronounced low-$T$ EEI correction to AHE was observed in the magnetic  semiconductor HgCr$_2$Se$_4$ ~\cite{HgCrSe}. A recent theoretical work  proposed that the low-$T$ EEI correction  could exits and anomalous Hall conductivity (AHC) should follow as $T^{1/2}/\ln(T_0/T)$ in three dimensional (3D)~\cite{Hg_Theory} material. Therefore, the effect of disorder on AHE remains inconclusive both theoretically and experimentally.
\par
\begin{figure*}
	\centering
	\includegraphics[width = 16 cm]{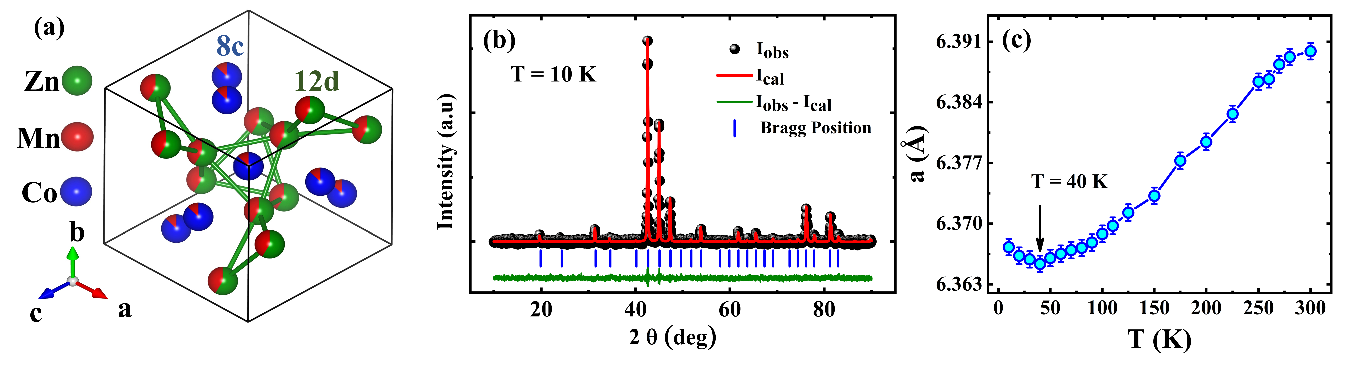}
	\caption{ (a) Crystal structure of Co$_7$Zn$_7$Mn$_6$) ({P4$_1$32}-right handed structure) as viewed along the [111] direction. (b) Powder x-ray diffraction pattern of Co$_7$Zn$_7$Mn$_6$ (data points) and Rietveld refinement curves (solid line) at 10 K and (c) Depict the thermal variation of lattice parameter(\emph{a}) } 
	\label{xrd}
\end{figure*}

In  the previous experimental works addressing the localization effect on AHE, the majority of the works are on systems having WL. There are only few experimental report where EEI is the primary cause of the localization~\cite{HgCrSe,SNKAUL,ZrVCoSn}. However, those few systems where EEI is prevalent, the temperature window where the upturn in $\rho_{xx}(T)$ is observed ({\it i.e.}, the region where EEI mediated localization dominates) is  narrow ($T_{min}$ is 30 K or less), and the upturn is rather weak. To examine the role of EEI in AHE, it is important to search for a system showing significant upturn in $\rho_{xx}(T)$ over a large $T$ range.  For the present study, we  chose $\beta$-Mn-type Co$_7$Zn$_7$Mn$_6$ alloy, which can be thought of being derived from Co$_{10}$Zn$_{10}$ by the substitution of Co and Zn by Mn. Below a critical temperature $T_c$$\sim$480 K, Co$_{10}$Zn$_{10}$ undergoes a transition from paramagnetic state to helimagnetic state~\cite{phaseD1,phaseD2} and $T_c$ decreases with the partial substitution of Mn. Just below $T_c$, it exhibits skyrmionic state in a small $T$ and field ($H$) (100 Oe $\lesssim H \lesssim$ 400 Oe) windows. At higher $H$ ($H>$ 1 kOe), the sample attains a completely ferromagnetic (FM) state. This alloy has a chiral cubic crystal structure (enantiomeric space group: $P4_132$ or $P4_332$, depending on its handedness) and the unit cell contain 20 atoms which are distributed over two Wyckoff sites [8$c$ and 12$d$, see Fig.~\ref{xrd} (a)]~\cite{natcom,phaseD2}. Previous studies reveal that 8$c$ sites are mainly occupied by Co atoms while Zn and Mn atoms prefer the hyper kagom\'e network of 12$d$ sites~\cite{occup1,glass1,glass2}. Due to multiple crystallographic sites and similar radii of Mn, Co and Zn atoms, antisite disorder is very likely to occur  in these $\beta$-Mn-type alloys~\cite{glass1,glass2}. These materials are ideal test bed for studying the disorder effect in AHE, because they show, (i) long range magnetic ordering with spontaneous magnetization, and have (ii) chiral structure lacking both inversion and mirror symmetry, which can lead to intrinsic Berry phase induced Hall effect. 
\par   
Till now Co-Zn-Mn alloys mostly studied to explore the skyrmionic state~\cite{method,phaseD2,SANS} via magnetization, ac susceptibility, small angle neutron scattering and transmission electron microscopy. However, little attention is given to their electronic transport properties. In fact previous Hall studies on Co-Zn-Mn gave some contradictory results. Zeng {\it et al.} reported the main mechanism behind the AHE to be skew scattering~\cite{trans2}, while Qi {\it et al.} found the dominance of   intrinsic mechanism~\cite{trans1}. In the present work, we have carefully investigated the AHE in the alloy Co$_7$Zn$_7$Mn$_6$ with the intention to see whether the effect of carrier localization  is affecting it. Our work indicates that the sample shows upturn in $\rho_{xx}$ due to EEI, while  AHE remains unaffected, which substantiate the theory proposed by Muttalib and W\"olfle, and Langenfeld and W\"olfle~\cite{EEI1,EEI2}.

   
\section{Experimental Details}
A polycrystalline sample of Co$_7$Zn$_7$Mn$_6$ was synthesized by the method as describe in previous report by Kosuke Karubel {\it et al.}~\cite{method}. The structural investigation of the sample was performed by powder x-ray diffraction (PXRD) in the $T$ range 10-300 K, using RIGAKU Smartlab (9KW) XG diffractometer fitted with a helium closed cycle refrigerator, using Cu K$\alpha$ radiation. The Rietveld refinement of the XRD data was performed using the MAUD software package~\cite{MAUD}. Magnetic measurements were carried out using the vibrating sample magnetometer module of a commercial physical properties measurement system (PPMS, Quantum Design) as well as on a SQUID-VSM (MPMS3) of Quantum Design. The standard four-probe technique was used to measure $\rho_{xx}$ on a cryogen-free high magnetic field  system (Cryogenic Ltd. UK) between 5 and 300 K. Hall measurements were perform using Physical Properties Measurement System (Quantum design Inc., USA) using four probe technique.


\section{Results}
\subsection{Powder X-ray diffraction}

 The Rietveld refinement of the temperature dependent PXRD data indicate that the sample retains its cubic structure with space group P$4_1$32, down to the lowest measured temperature of 10 K [Fig.~\ref{xrd} (b)]. Fig. \ref{xrd} (c) shows the thermal variation of lattice parameter $a$ (in \AA). The cubic lattice parameter $a$ decreases monotonically down to 40 K with lowering of $T$. Interestingly, below  40 K, we notice a clear anomaly as $a$ increases with decreasing $T$. As a result, we find a minimum at around 40 K in the $a$ versus $T$ data, and the data below 40 K show a negative thermal expansion ($da/dT <$0).

\subsection{Magnetization} 

\begin{figure*}[t]
	\centering
	\includegraphics[width = 16cm]{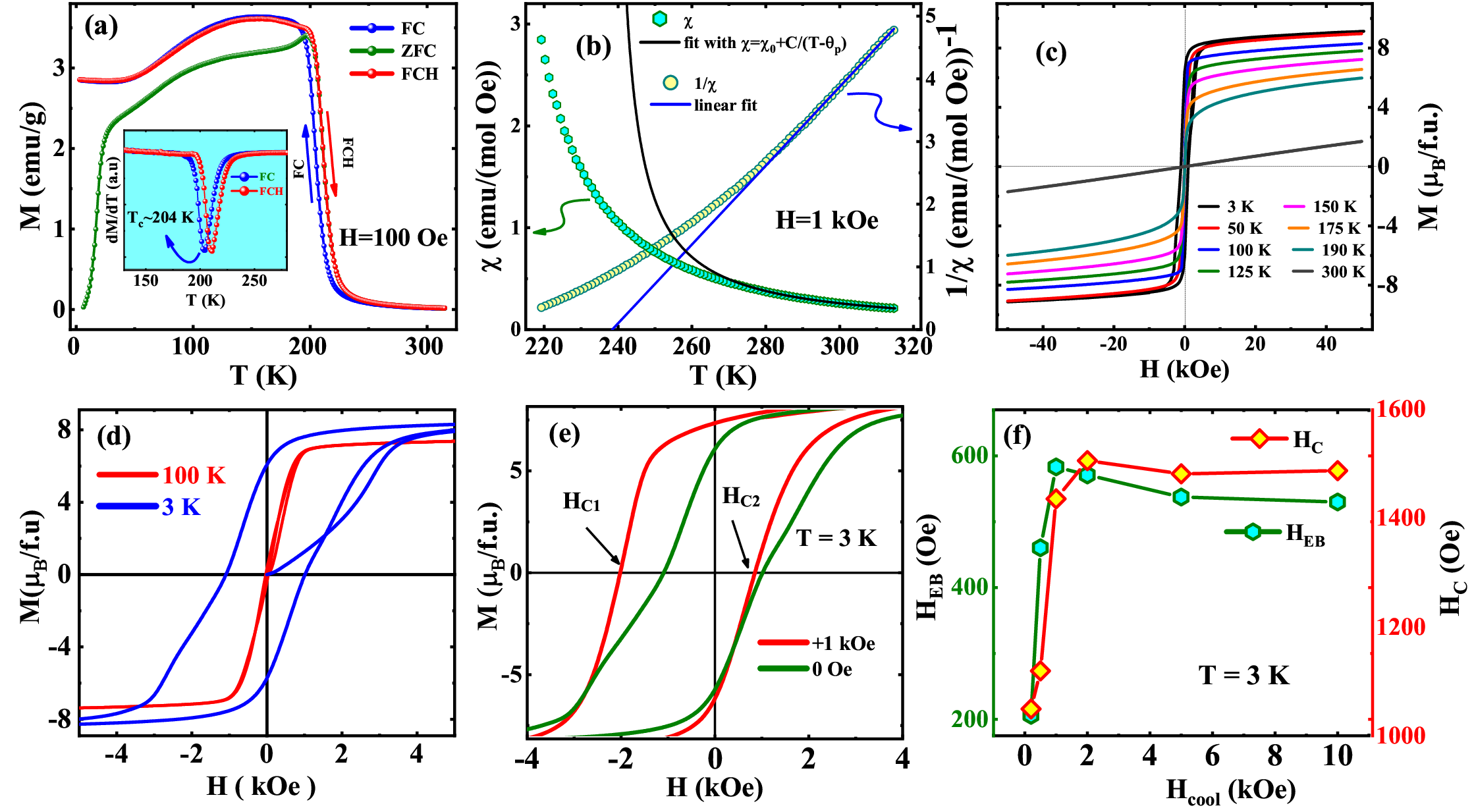}
	\caption {(a) The magnetisation (M) as a function of the temperature (T) measured in zero field-cooled-heating (ZFCH), field-cooled (FC) and field-cooled-heating (FCH) protocols under H = 100 Oe. The inset shows the temperature derivative of $M(T)$. (b) Susceptibility ($\chi$) versus temperature ($T$) fitted with modified Curie-Weiss law (left axis) and inverse susceptibility ($\chi$$^{-1}$) versus temperature ($T$) data with liner fitting (right axis), measured at H = 1 kOe. (c)  Isothermal magnetization at different temperatures measured between $\pm$50 kOe. (d) Enlarge view of the $M$ vs $H$ curve at 3 K and 100 K plotted between $\pm$ 5 kOe. (e) Shows isothermal magnetization data measured at 3K after being cooled at zero and +1 k Oe. (f) Shows the variation of exchange bias field (\emph{H$_{EB}$}) on left axis and coercive field (\emph{H$_C$}) on right axis as a function of cooling field (\emph{H$_{cool}$}) at 3K. }
	\label{Magnetic}
\end{figure*}

Fig.~\ref{Magnetic} (a) shows the $T$ variation of magnetization ($M$) in an externally applied magnetic field of $H$ = 100 Oe. The measurements were performed in zero-field-cooled-heating (ZFC), field-cooling (FC) and field-cooled-heating (FCH) protocols. The $T$-dependence of $dM/dT$ for FC and FCH measurements is shown in the inset of Fig.~\ref{Magnetic} (a). From the minimum of the $dM/dT$ vs $T$ plot, the transition from paramagnetic state to a helimagnetic state~\cite{phaseD1,phaseD2,heli} is found to be at $T_c$$\sim$204 K. Interestingly, around $T_c$, there is a clear thermal hysteresis between FC and FCH data which indicates a first order like magnetic transition. The hysteresis is present for different temperature ramping rates (5 K and 10 K per min) and also for different magnetometers (PPMS and SQUID-VSM). However,  $T$ dependent PXRD does not show any major anomaly in $a$ around $T_c$ [see Fig.~\ref{xrd} (c)], and it rules out any structural change at $T_c$.  A sharp drop of $M$ below 35 K is observed in the ZFC data, which matches well with the onset of reentrant spinglass state reported previously based on ac susceptibility measurements~\cite{glass1,glass2}. The spin glass transition is likely due to the geometrical frustration, which is intrinsic to $\beta$-Mn type structure, and disorder present in the system~\cite{method}. Our $T$ dependent PXRD [see Fig.~\ref{xrd} (c)] result shows an anomaly in lattice parameter at below 40 K, which is close to the spin-glass freezing temperature. This may indicate a close interplay between magnetic and structural aspects in the sample. 
\par
We have examined the  $\chi$$^{-1}$ vs $T$ data measured at $H$= 1 kOe between 220 to 315 K, and it is shown in the right panel of Fig.~\ref{Magnetic} (b). $\chi$$^{-1}$ is seen to deviate from linearity below 285 K and from linear fit above 285 K, we obtain Weiss temperature ($\theta_p$) to be 238 K, which is quite high as compared to $T_c$. We also tried to fit the $\chi$ vs $T$ data with modified Curie-Weiss law: $\chi$ = $\chi_0 + C/(T-\theta_p)$, where $\chi_0$ is a $T$ independent term, $C$ is the Curie constant and $\theta_p$ is the Weiss temperature. As we can see from the left panel of Fig.~\ref{Magnetic} (b), experimental data start to deviate from the fitting below 280 K which indicates that the sample does not obey the Curie-Weiss law. Co$_7$Zn$_7$Mn$_6$ is an itinerant magnetic system and a deviation from the Curie-Weiss law may occur~\cite{itinerant}.  

\par
The isothermal $M$ vs $H$ curves recorded at different constant temperatures ($T =$ 3 K, 50 K, 100 K, 125 K, 150 K, 175 K, 190 K, and 300 K), between $\pm$50 kOe, are plotted in Fig.~\ref{Magnetic} (c). $M$ shows fairly saturating behavior with $H$ below $T_c$. The saturation magnetization ($M_s$) takes maximum value 9.18 $\mu_B$/f.u. at 3 K. It is evident [see Fig.~\ref{Magnetic} (c)] that the sample shows finite coercive field ($H_{C} \sim$ 1 kOe) only at 3 K, while $H_{C}$ is negligible at 100 K data. This is consistent with the previous report that hysteresis loop is only observed below the re-entrant spin glass transition temperature~\cite{glass1}. 

\par
Interestingly, we observe a virgin loop effect~\cite{sbroy1,sbroy2} at 3 K, where the virgin line lies outside the hysteresis loop [see Fig.~\ref{Magnetic} (d)]. Such virgin loop effect often occurs due to the presence of a field induced arrested state. The sample was earlier found to have a spin glass-like state at low temperature, which can be responsible for the virgin loop effect~\cite{loop1,loop2}.

                                              
\par
Several  spin-glass systems were  found to show exchange bias (EB) effect~\cite{giri,EB1,EB2,EB3}, which is  characterized by the  horizontal shift (along the $H$ axis) of the isothermal $M-H$ curve when cooled under a field ($H_{cool}$) from above the magnetic transition temperature. The values of EB and $H_C$ can be defined as $H_{EB} = -(H_{c1} + H_{c2})/2$ and $H_C = |H_{c1} - H_{c2}|/2$ , where $H_{c1}$ and $H_{c2}$ denote the negative and positive fields at which $M$ turns zero, respectively~\cite{EB4,EB5}. Fig.~\ref{Magnetic} (e) shows the $M-H$ loops at 3 K both in the ZFC ($H_{cool}$ = 0) and FC ($H_{cool}$ = 1 kOe) conditions. Although we recorded the data  for $H = \pm$ 50 kOe, an enlarged view in the region $\pm$4 kOe are shown for a  better clarity. Clearly, $M-H$  loop shift asymmetrically along the field axis in the direction opposite to $H_{cool}$, indicating the presence of finite EB in the system. We recorded the $M-H$ loops for different values of $H_{cool}$, and the variation of $H_{EB}$ and $H_C$ with $H_{cool}$ is shown in Fig.~\ref{Magnetic} (f).  $H_{EB}$  initially rises sharply  with increasing $H_{cool}$ up to 1 kOe followed by sluggish decrease on higher fields. The maximum value of $H_{EB}$ is found to be 580 Oe for $H_{cool}$ = 1 kOe at 3 K.  The antisite disorder plays an important role towards the observed EB, because it can give rise to varied magnetic interactions depending upon the metal ion and its position in the lattice~\cite{MnPtGa}. $H_C$ follows $H_{cool}$ variation similar to $H_{EB}$.

\subsection{Electrical Resistivity}
\label{section:resistivity}
\begin{figure}[h]
	\centering
	\includegraphics[width = 8 cm]{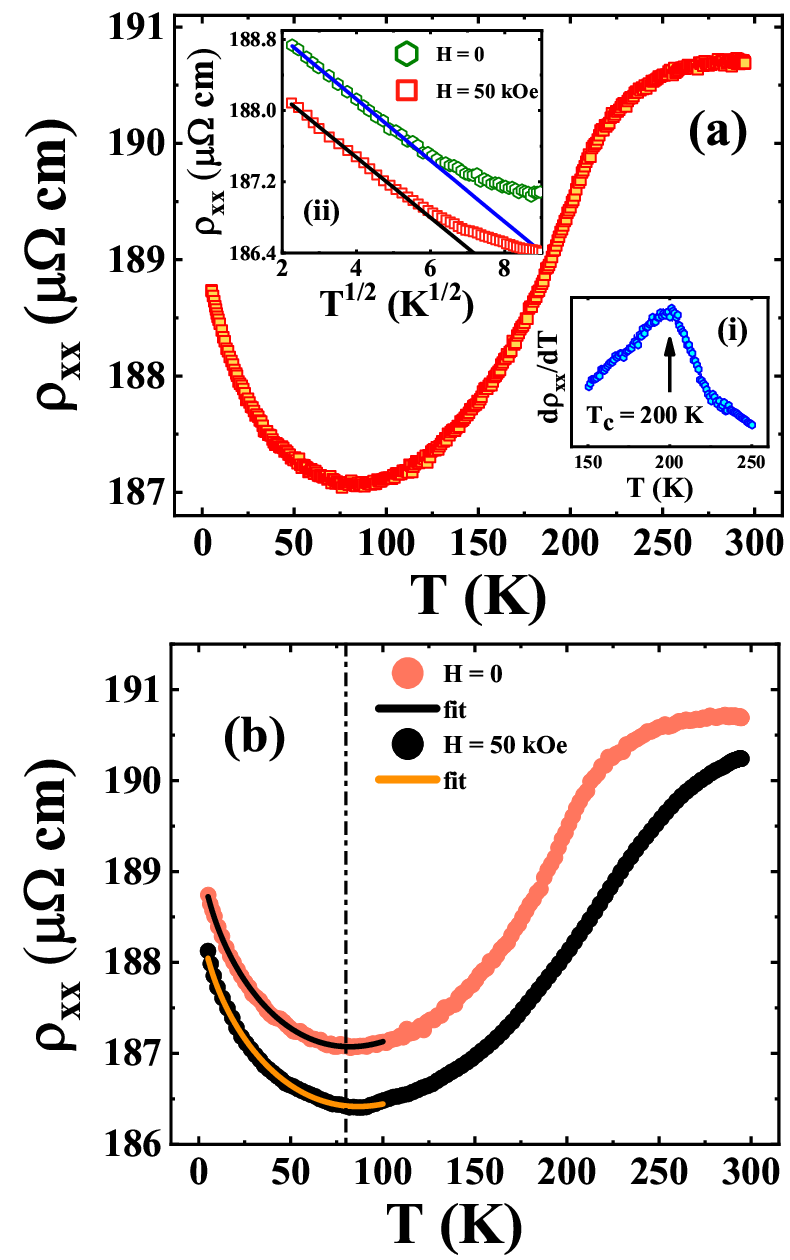}
	\caption{ (a) shows the $T$ variation of resistivity ($\rho_{xx}$). The inset a(i) shows the $\frac{d{\rho_{xx}}}{d{T}}$ vs $T$ curve and a(ii) shows $\rho_{xx}$ vs $T^{1/2}$ plot below 50 K, along with a linear fit to the data. (b) shows $\rho_{xx}$ vs $T$ data at $H$= 0, 50 kOe fitted with eqn.~\ref{eqn:rho1} }.
	\label{Resistivity}
\end{figure}

  We have shown the $T$ variation of zero-field resistivity [$\rho_{xx}$($T,H=0$)] in the main panel of Fig.~\ref{Resistivity} (a) for Co$_7$Zn$_7$Mn$_6$ in the range $5K\le T\le300K$. $\rho_{xx}$ vs $T$ data  show a typical metallic behavior ($\frac{d{\rho_{xx}}}{d{T}}>0$) in the high-$T$ region. However, below 75 K, $\rho_{xx}$ starts to rise with decreasing $T$ giving rise to a resistivity minimum at $T_{min}$ = 75 K. There is a change in slope at $T_c$ $\sim$ 200 K [see  the peak in the $d\rho_{xx}/dT$ curve in the inset (i) of Fig.~\ref{Resistivity} (a)], which is close to the magnetic transition temperature (= 204 K). 
\par
The low-$T$ rise can have multiple origins, such as Kondo effect, WL or EEI ~\cite{RMP}. Kondo effect involves $\log (T)$ upturn of $\rho_{xx}(T)$ at low temperature~\cite{kon1,kon2,kon3}, which is absent in our data, and it rules out the Kondo type localization of charge carriers. The contribution from WL  to $\rho_{xx}$ varies as $T^{p/2}$ ($p$ = 3/2, 2 or 3) in 3D disordered system~\cite{RMP}, and it is very sensitive to $H$. In case of La$_{1-x}$A$_x$MnO$_3$ (A=Ca,Sr,Ba or Pb), showing resistivity upturn due to WL, the minimum in $\rho_{xx}$ shifts to lower $T$ on application of $H$ and it nearly vanishes for higher applied $H$~\cite{manganties1,manganties2}.
\par
Altshuler and Aronov have shown that EEI gives rise to increase in $\rho_{xx}$  with  decreasing $T$ instead of usual metallic behavior~\cite{Aronov}. In presence of disorder, the relative change in $\rho_{xx}$ due to EEI is estimated to be 
\begin{equation}
	\delta \rho_{xx} = \left [\frac{\rho_0 - \rho_{xx}(T)}{\rho_0}\right]  \propto \frac{\sqrt{T\tau_e}}{(P_Fl_e)^2}
	\label{EEI}
\end{equation}
Here,  $\tau_e$ is the characteristic mean free time, $P_F$ is Fermi momentum and $l_e$ is the mean free path between two successive collisions. This expression is valid for $K_BT$ $\ll$ $\hbar$/$\tau_e$, and this inequality is satisfied for $T<T_{min}$. The true cause of this upturn below $T_{min}$ is closely associated with the quantum interference effect in presence of disorder~\cite{sr_weak,Aronov,Aronov2}.
\par

Fig.~\ref{Resistivity} (b) shows the resistivity minimum of Co$_7$Zn$_7$Mn$_6$ which does not shift with $H$. On careful examination of the low-$T$ part of $\rho_{xx}$,  we find that  it follows a $T^{1/2}$ dependence [see inset of Fig.~\ref{Resistivity} (a) (ii)] for both  $H$ = 0 and  50 kOe data below about 35 K. This indicates  that the EEI plays the dominant role towards the low-$T$ upturn~\cite{sr_weak,prb_weak}. We observe that $\rho_{xx}$($T$, $H$ = 50 kOe) lies below the zero field counterpart due to the small but finite negative magnetoresistance. 
 
\par
Considering all possible contributions to resistivity in our system, we model the $T$ variation of $\rho_{xx}$ below $T_c$ as: 

\begin{equation}
	\begin{array}{l}
	\rho_{xx}(T,H) = \rho_{xx0} - \gamma_{EEI}{T^{1/2}} + \beta_{e-m}T^2 \\ 
	        \qquad \qquad \quad	+ \alpha_{e-p}(T/{\theta_D})^5\int_{0}^{\theta_D/T}\frac{x^5dx}{(e^x-1)(1-e^{-x})}
	\end{array}
\label{eqn:rho1}
\end{equation}

\noindent In eqn.~\ref{eqn:rho1},  $\rho_{xx0}$ is a temperature independent term, while the second and the third terms respectively represent the electron-electron (e-e) interaction and electron-magnon (e-m) scattering contribution. The last term represents the electron-phonon (e-p) scattering contribution according to Bloch-Gr\"uneisen model ($\theta_D$ is the Debye temperature)~\cite{thetaD}. The continuous lines in Fig.~\ref{Resistivity} (b) indicate the fitting to the data by eqn.~\ref{eqn:rho1} for both $H$ = 0 and 50 kOe curves below 100 K. The fitting parameters are provided in table~\ref{table:rho}. We choose  $\theta_D$ to be 320 K for both the values of $H$. The coefficient of electron-magnon scattering term ($\beta_{e-m}$) decreases under $H$, because an applied field reduces the spin-disorder scattering.  

\begin{table}
	\centering
	\begin{tabular}{|c |c |c |c |c |c |}
		\hline
		$H$~ &   $\rho_{xx0}$  & $\gamma_{EEI}$  &  $\beta_{e-m}$  &   $\alpha_{e-p}$  \\
		(kOe) &   ($\mu\Omega$ cm)  &  ($\mu\Omega$ cm K$^{-1/2}$)  &  ($\mu\Omega$ cm K$^{-2}$)  &    ($\mu\Omega$ cm)  \\
		\hline
			\hline
		0 & 189.54(2)    &  0.36(3)   & 11.7(7)$\times$10$^{-5}$    & 0.83(3)      \\
		\hline
		50 & 188.81(3)    &  0.34(3)   & 10.2(3)$\times$10$^{-5}$  & 0.82(8)       \\
		\hline
	\end{tabular}
	\caption{The fitting parameters obtained by fitting the resistivity data using eqn.~\ref{eqn:rho1}}
	\label{table:rho}
\end{table} 
  
\par
\begin{figure}[h]
	\centering
	\includegraphics[width = 8 cm, height= 10 cm]{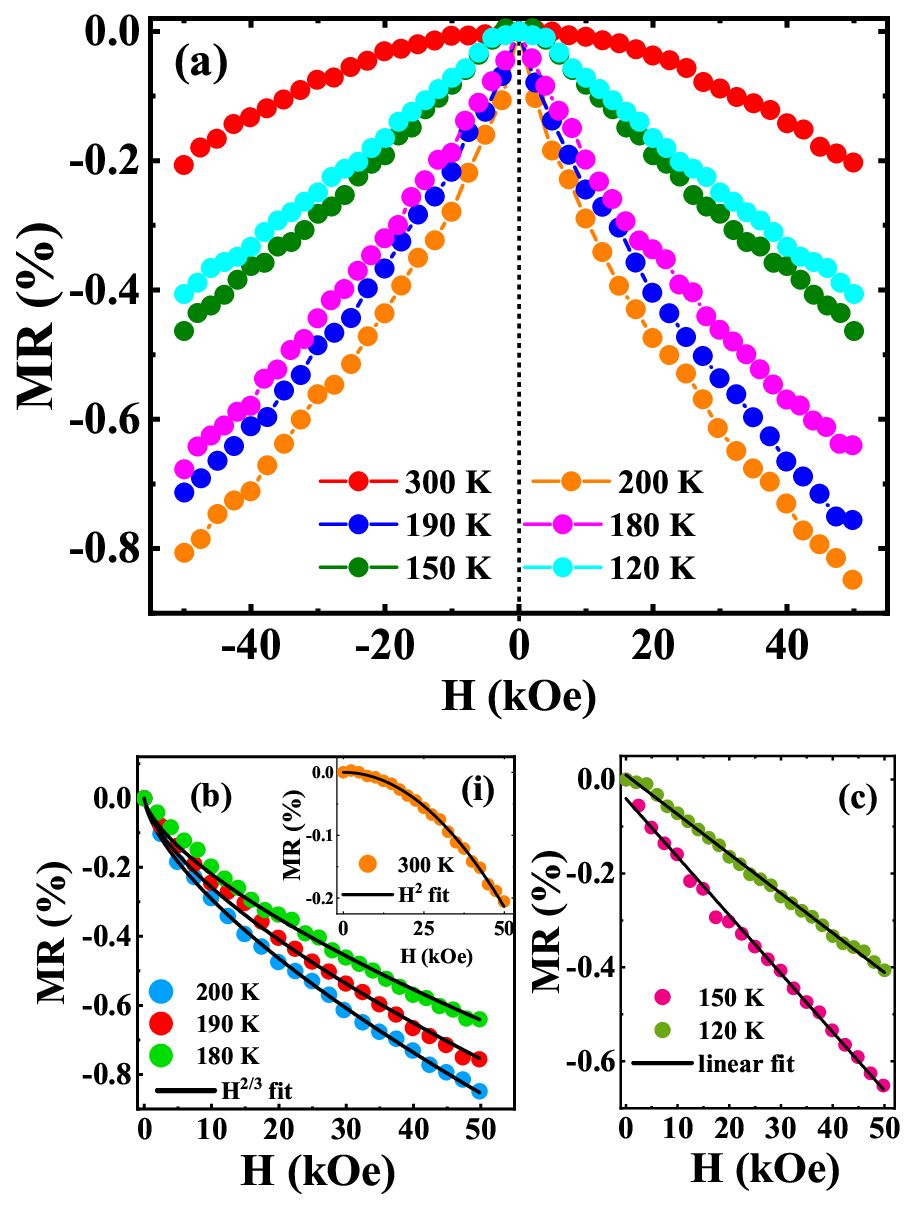}
	\caption{ (a) shows isothermal field dependent MR . (b) main panel shows $H^{2/3}$ dependence of $MR$ vs $H$ near the transition temperature, inset shows the $H^2$ dependence of $MR$ vs $H$  at 300 K. (c) shows the linear field dependence of MR well below the transition temperature.}
	\label{magres}
\end{figure}

We  also studied the isothermal $H$ variation of $\rho_{xx}$  at different $T$. The field variation of magnetoresistance (MR) [$= (\rho_{xx}(H) - \rho_{xx}(0))/(\rho_{xx}(0)$)] is shown in Fig.~\ref{magres} (a). At 300 K, MR is found to obey an $H^2$ dependence [see inset of Fig. ~\ref{magres} (b)], while MR shows a $H^{2/3}$ variation around $T_c$ [see main panel of ~\ref{magres} (b)]. Well below $T_c$ (150, 120 K), a linear variation of MR with $H$ is found [see Fig. ~\ref{magres} (c)]. It is generally believed that for a 3$d$ transition metal based  intermetallic alloy showing FM-like order, the magnetic contribution to the resistivity arises due to the scattering of the delocalized $s$ electrons with the partially localized 3$d$ electrons. Under the application of $H$, the spin dependent scattering diminishes leading to negative MR. 
\par
Although the value of MR is quite small, its field variation in different $T$ range is quite interesting. There are several theoretical works addressing the $H$ dependence of MR in the transition metal based itinerant magnets~\cite{sd1,sd2}, and the theory predicts an $H^{2/3}$ variation around $T_c$~\cite{sd2}, which is also the case for our present sample. A linear $H$ variation of MR is predicted in the $s-d$ scattering model, which is clearly observed here. The high temperature (well above $T_c$) $H^2$ variation of MR in the $s-d$ model is also evident in our 300 K data. It is clear that the observed small negative MR in the present compound arises due to the suppression of $s-d$ scattering by the applied field, and  Co$_7$Zn$_7$Mn$_6$ is turned out to be classic example where the prevailing theory of MR works well at least up to 50 kOe.        

\subsection{Hall measurements}
\begin{figure}[t]
	\centering
	\includegraphics[width = 8.5 cm]{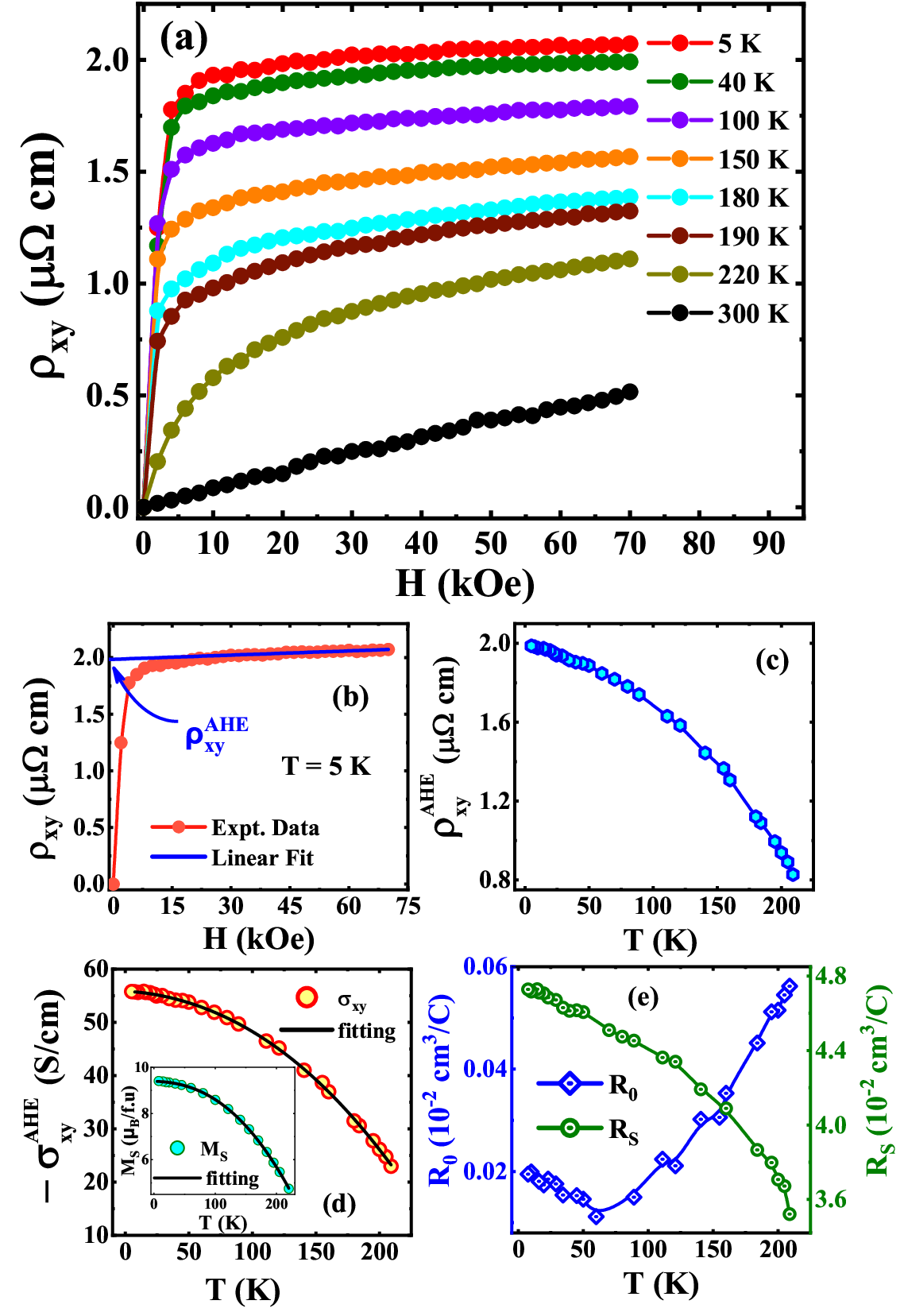}
	\caption{ (a) Hall resistivity ($\rho_{xy}$) vs $H$ at different constant $T$. (b) shows $\rho_{xy}$ vs $H$ at $T$ = 5K with a linear fit to the high field data. ($\rho_{xy}$) vs $H$, $\rho_{xy}^{AHE}/\rho_{xx}$ vs $\rho_{xx}$ plot with linear fit. (c) $T$ variation of $\rho_{xy}^{AHE}$. (d) $\sigma_{xy}^{AHE}$ vs $T$ fitted with eqn.~\ref{eqn:bloch}  and the inset shows the saturation magnetization ($M_s$) as a function of $T$ and fitted with eqn.~\ref{eqn:bloch}. (e) $T$ variation of ordinary Hall  coefficient ($R_0$) and anomalous Hall coefficient ($R_s$).}
	\label{Hall}
\end{figure}

 The major outcome of the present work is based on the Hall effect study of Co$_7$Zn$_7$Mn$_6$.
 The Hall resistivity ($\rho_{xy}$) data  as a function of $H$ at various $T$ are  shown in Fig.~\ref{Hall}(a). Here, the current is allowed to flow along the $x$ axis, while the Hall voltage was measured along the $y$ direction with the magnetic field along the $z$ axis. We measured $\rho_{xy}$ for both positive and negative values of $H$, and the final $\rho_{xy}$ versus $H$ plot was drawn using the formula $\rho_{xy}(H) = \frac{1}{2}[\rho_{xy}(+H) - \rho_{xy}(-H)]$. This helps us to eliminate any MR contribution in the data. The $\rho_{xy}(H)$ is found to be highly nonlinear below about 200 K indicating the presence of AHE. There is a sharp increase in the low field region ($H$= 0 $\sim$ 5 kOe), and the data tend to saturate at higher fields with small positive slope. This sluggish increase in $\rho_{xy}$ at higher $H$ is due to the contribution from ordinary Hall effect (OHE). $\rho_{xy}(H)$  mimics the $M$ versus $H$ data indicating the influence of $M$ towards the Hall voltage. Notably, $\rho_{xy}(H)$ is linear at 300 K (well above $T_c$), because  AHE vanishes in the paramagnetic state.
\par
Typically in a magnetic material with magnetization $M$, $\rho_{xy}$ can be expressed as (in cgs unit) 
\begin{equation}
\rho_{xy} = \rho_{xy}^{OHE} +  \rho_{xy}^{AHE} = R_0H + 4\pi R_S M	
	\label{eqn:Hall}
\end{equation}
 
\noindent where, $\rho_{xy}^{OHE}$ and $\rho_{xy}^{AHE}$ are the ordinary and anomalous Hall resistivities with coefficients $R_{0}$ and $R_S$ respectively. By linear fitting the high field region ($H\geq$ 30 kOe) of the $\rho_{xy}$ vs $H$ curve [Fig.~\ref{Hall}(b)], we obtain $\rho_{xy}^{AHE}$ and $R_0$ from the $y$-intercept and the slope, respectively. The $\rho_{xy}^{AHE}$ is found to decrease monotonously with increasing temperature [see Fig.~\ref{Hall}(c)]. The anomalous Hall conductivity ($\sigma_{xy}^{AHE}$) is calculated using the formula $\sigma_{xy}^{AHE} = -\rho_{xy}^{AHE}/[(\rho_{xy}^{AHE})^2+(\rho_{xx})^2]$ and its temperature dependence is shown in Fig.~\ref{Hall}(d). 

\par
$M_s$ vs $T$ curve [see inset of Fig.~\ref{Hall} (d)] is found to obey the well-known spin-wave (SW) equation~\cite{Bloch1,Bloch2} 
\begin{equation}
	M_s(T) = M_s(0)(1-AT^{3/2}-BT^{5/2})	
	\label{eqn:bloch}
\end{equation}

\noindent Here $M_s(0)$ is the value at 0 K. The fitted values of the parameters $A$ and $B$ are found to be 2.928$\times$10$^{-5}$ K$^{-3/2}$ and 5.531$\times$10$^{-7}$ K$^{-5/2}$ respectively. Interestingly, $\sigma_{xy}^{AHE}$  follows a similar equation that of eqn.~\ref{eqn:bloch} [i.e $\sigma_{xy}^{AHE}$ $\propto$ $(1-AT^{3/2}-BT^{5/2})$] down to 5 K without any visible anomaly below $T_{min}$ and the corresponding fitted parameters $A$ and $B$ are 8.25$\times$10$^{-5}$ K$^{-3/2}$ and 5.292$\times$10$^{-7}$ K$^{-5/2}$ respectively. The absence of any anomaly below $T_{min}$ indicates that both $\sigma_{xy}^{AHE}$ and $\rho_{xy}^{AHE}$ remain unaffected by the e-e localization effect present in $\rho_{xx}$ within the accuracy of our measurements. To check the result, we measured Hall voltage using two different instruments (from Quantum Design, USA and Cryogenic Ltd, UK). However, data from both the measurements are identical providing no signature of EEI to AHE. This finding provides an  experimental evidence to the theoretical prediction~\cite{EEI1,EEI2} that the correction to $\sigma_{xy}^{AHE}$ is identically zero  even though there is a finite low-$T$ upturn in the $\rho_{xx}(T)$ data due to EEI. 
\par
Fig.~\ref{Hall}(e) shows the temperature variation of $R_{0}$ and $R_{S}$ calculated using the formula $R_{S} =\rho_{xy}^{AHE}/(4\pi M_S$). It is clearly seen that the localization effect is present in $R_{0}$ at low temperature but not in $R_{S}$. $R_{S}$ is found to be two order of magnitude greater than the $R_0$, which indicates the dominance of anomalous Hall resistivity. The coefficient  $R_0$ is found to be positive, which indicates holes as  majority charge carriers. The carrier concentration ($n_h$) is calculated using $n_h$ = 1/($R_0$e), and  it is found to be 3.2$\times$10$^{22}$ cm$^{-3}$ at 5 K.

\section{Discussion}
 
The most important observation from the present study is  associated with the electronic transport study of Co$_7$Zn$_7$Mn$_6$. We find a robust upturn in the longitudinal resistivity versus temperature  data below about 75 K, which remains almost unaffected even under 50 kOe of field. Such low temperature upturn in otherwise metallic alloy is generally attributed to localization of charge carriers. Our careful investigation of zero field and with field $\rho_{xx}(T)$ data show that at low temperature $\rho_{xx}(T)$ follow $T^{1/2}$ dependence [see inset of Fig.~\ref{Resistivity} (a)], which confirms that the electron-electron interaction  is primarily responsible for this  low-$T$ upturn~\cite{RMP}. Evidently, this EEI mechanism is insensitive to the magnetic field (at least for $H \leq$ 50 kOe).

\par
The signature of EEI is also seen in the regular Hall coefficient, $R_0$. It shows an upturn below about 70 K, which is similar to the upturn in $\rho_{xx}(T)$. A rise in $R_0$ also indicates a decrease in free charge carriers as $R_0$ = 1/($ne$), where $n$ is the carrier concentration and $e$ is the electronic charge.  Interestingly, the localization effect observed in $\rho_{xx}(T)$ and $R_0$  at low temperature is completely absent in all components of AHE, \emph{i.e,} in $\rho_{xy}^{AHE}(T)$, $R_s(T)$ and even in $\sigma_{xy}^{AHE}(T)$. As evident from Fig.~\ref{Hall}(d),  $\sigma_{xy}^{AHE}(T)$ varies monotonously and it obeys the SW equation (eqn.~\ref{eqn:bloch}). The fit with SW law does not show any deviation at low temperature, particularly below 75 K, where EEI localization in  $\rho_{xx}$ and $R_0$ is evident. This result supports the previous theoretical prediction that EEI correction to AHE identically vanishes for both skew scattering and side-jump mechanisms~\cite{EEI1,EEI2}. It is to be noted that the change in the carrier concentration due to localization, as obtained from ordinary Hall coefficient, is 5.57$\times$10$^{22}$ (at 60 K) to 3.2$\times$10$^{22}$ (at 5 K). Such change may not affect the electronic energy band structure required for change in anomalous Hall coefficient.

\par
Recently Yang {\it et al.}  showed the presence of EEI contribution towards the AHE in the  semiconducting HgCr$_2$Se$_4$ single crystal with  a $T^{1/2}$ dependence of $\sigma_{xy}^{AHE}(T)$ at sub Kelvin temperature~\cite{HgCrSe}. The compound shows low-$T$ electron localization due to EEI with  $\rho_{xx}(T) \propto T^{1/2}$. The theories that rule out the contribution of EEI to AHE~\cite{EEI1,EEI2} assume (i) the sample to have the mirror symmetry in its crystal structure, and (ii) take care only the extrinsic mechanisms (side jump and skew scattering) ignoring the Berry phase induced intrinsic mechanism. Yang {\it et al.} argued that although  HgCr$_2$Se$_4$ possess overall mirror symmetry, it may be broken locally at the site of the disorder. In addition, intrinsic contribution of Hall effect may also play a role towards the observed localization effect in $\sigma_{xy}^{AHE}(T)$. The effect of EEI in $\sigma_{xy}^{AHE}(T)$ is further substantiated by a recent theoretical work~\cite{Hg_Theory}, which  proposed that $\sigma_{xy}^{AHE}(T)$ should follow $T^{1/2}/\ln(T_0/T)$ (for 3D systems) type $T$ dependence due to EEI effect at low-$T$. This theory takes into account the contribution from  Cooper channel for the  purely repulsive interaction in a non-superconducting metal, which was overlooked in the previous studies.  HgCr$_2$Se$_4$ system has relatively lower carrier density ($\sim$ 10$^{15}$-10$^{18}$ cm$^{-3}$)  and  $\rho_{xx} \sim$  10$^{-2}$~$\Omega$ cm, whereas in Co$_7$Zn$_7$Mn$_6$, the carrier density is found to  10$^{22}$ cm$^{-3}$ and $\rho_{xx}\sim$  2$\times$ 10$^{-5}~\Omega$ cm. The low carrier density in HgCr$_2$Se$_4$ may have an influence on the observed correction in $\sigma_{xy}^{AHE}(T)$. Notably, the effect of carrier localization in AHE is also absent in Co$_2$FeSi thin films~\cite{SNKAUL} and Zr$_{1-x}$V$_x$Co$_{1.6}$Sn semimetal~\cite{ZrVCoSn}, which are otherwise metallic with large carrier density.  

\par
Though there is a lattice anomaly at around 40 K, it is not reflected in our Hall conductivity data. The change in $a$ is only 0.03\%, which possibly too weak to provide any detectable signature in our Hall data.
\par
In conclusion, we fail to observe the effect of correlation induced electron localization in the anomalous Hall effect in the chiral compound Co$_7$Zn$_7$Mn$_6$. Although this result is consistent with the theoretical models proposed by the group of P. W\"olfle ~\cite{EEI1,EEI2}, it is in sharp contrast with recent experimental and theoretical works~\cite{HgCrSe,Hg_Theory}. The most of the theories proposed so far consider the presence of mirror symmetry in the lattice and primarily concentrated on extrinsic mechanisms. Therefore, it is important to address the issue with theoretical models where the system has broken mirror symmetry (such as the present Co$_7$Zn$_7$Mn$_6$) and with the inclusion of intrinsic contribution towards AHE.  

\section*{Acknowledgments}
PC wishes to thank DST-INSPIRE program for the research assistance. The UGC DAE-CSR Kolkata center, where the low-temperature magneto-resistance measurements were performed, is duly acknowledged. We thank Prof. Chandan Mazumdar, SINP, Kolkata for Hall measurements.

\end{document}